\documentclass[prl,floats,twocolumn]{revtex4-1}
\usepackage{amsmath}
\usepackage{bbm}
\usepackage{graphicx}
\usepackage{color}

\begin{document}

\title{Competition between disorder and interaction effects in 3D Weyl semimetals}
\author{J. Gonz\'alez}
\address{Instituto de Estructura de la Materia,
        Consejo Superior de Investigaciones Cient\'{\i}ficas, Serrano 123,
        28006 Madrid, Spain}

\date{\today}

\begin{abstract}
We investigate the low-energy scaling behavior of an interacting 3D Weyl semimetal in the presence of disorder. In order to achieve a renormalization group analysis of the theory, we focus on the effects of a short-ranged-correlated disorder potential, checking nevertheless that this choice is not essential to locate the different phases of the Weyl semimetal. We show that there is a line of fixed-points in the renormalization group flow of the interacting theory, corresponding to the disorder-driven transition to a diffusive metal phase. Along that boundary, the critical disorder strength undergoes a strong increase with respect to the noninteracting theory, as a consequence of the unconventional screening of the Coulomb and disorder-induced interactions. A complementary resolution of the Schwinger-Dyson equations allows us to determine the full phase diagram of the system, showing the prevalence of a renormalized semimetallic phase in the regime of intermediate interaction strength, and adjacent to the non-Fermi liquid phase characteristic of the strong interaction regime of 3D Weyl semimetals.
\end{abstract}

\maketitle



{\em Introduction.---}
During more than a decade we have witnessed the discovery of a number of materials whose electronic properties have been defeating the conventional description of solid state systems. Starting with graphene\cite{novo} and ending with
the 3D Weyl semimetals\cite{w1,w2}, these materials display degenerate bands that touch only at isolated points in momentum space, with a linear dispersion similar to that of relativistic particles. 
This introduces novel topological features in the description of the electron systems\cite{kar,franz,bur3,son,bur2,ong,vish,ran,bur,zang,krem,dai,has}, which has important consequences for the transport properties of the materials.

The Coulomb interaction has to play also an important role in those systems, as it remains long-ranged at the nodal points. A key property is the scale dependence of the quasiparticle parameters at low energies, already observed in the case of the Fermi velocity of graphene\cite{exp}. Regarding the 3D semimetals, similar effects have to exist implying the renormalization of the Fermi velocity\cite{hosu,lewk} as well as of the electron quasiparticle weight. It has been shown that, for sufficiently large number of nodal points, the renormalization of the latter should be the dominant effect at low energies, with the potential to drive the system to a non-Fermi liquid phase\cite{rc,hep}.

In this picture, it would be convenient to assess the impact of disorder in the electron system. It was found long ago that, in the presence of a random disorder potential, 3D semimetals may undergo a transition to a phase characterized by developing a nonvanishing density of states at the nodal points\cite{fra}. Recently, there has been much interest in understanding the nature of that transition\cite{mura,gos,nomu,herb,brou,kosh,nand,radz,das,das2,roy,das3,carp,juri,syz,fed}. It remains unknown to a large extent, however, whether the disorder can modulate the interaction effects in such semimetals or, vice versa, whether the long-range Coulomb interaction can modify the disorder-driven transition.

In this paper, we investigate the low-energy scaling behavior of an interacting 3D Weyl semimetal, when it is under the effect of a random disorder potential. For the sake of achieving a renormalization group analysis of the interacting theory, we will choose a disorder potential with suitable short-range correlations, though we will see that this choice is not essential to locate the different phases of the Weyl semimetal. We will show that the most important feature is the unconventional screening of the Coulomb and the disorder-induced interactions, implying a strong increase of the critical disorder strength for the disorder-driven transition. This will determine the shape of the phase diagram of the system, leading to the prevalence of a renormalized semimetallic phase in the whole regime of intermediate interaction strength, and adjacent to the non-Fermi liquid phase characteristic of the strong interaction regime of 3D Weyl semimetals.

{\em Large-$N$ renormalization group approach.---}
Our starting point is a model of $2 N$ two-component Weyl spinors $\{ \psi_i \}$ with long-range Coulomb interaction, governed by the action
\begin{eqnarray}
S   & = &   \int d^3 r dt \; \psi^{\dagger}_i({\bf r})  \left( - i \partial_t
     - i v_F \gamma_0 \mbox{\boldmath $\gamma $}  \cdot  \mbox{\boldmath $\partial $} 
       - e_0  \phi ({\bf r})   \right)  \psi_i ({\bf r})          \nonumber        \\        
    &   &    +  \int d^3 r \; \psi^{\dagger}_i({\bf r})  \psi_i ({\bf r}) \: \eta ({\bf r})
\label{act}
\end{eqnarray}
where $\eta ({\bf r})$ is the field representing the disorder, $\phi ({\bf r})$ stands for the scalar interaction potential, and $\{ \gamma_\alpha \}$ is a set of matrices satisfying $\{\gamma_\alpha , \gamma_\beta \} = 2g_{\alpha \beta}$ \cite{metr}. Assuming that the Fermi velocity $v_F$ is typically much smaller than the speed of light, we can neglect retardation effects in the $e$-$e$ interaction and take the free propagator of the scalar potential in momentum space as $D_0({\bf q},\omega ) = 1 /{\bf q}^2$. On the other hand,     
$\eta ({\bf r})$ corresponds in general to a random potential with zero average and a variance
\begin{equation}
\overline{\eta ({\bf r}) \eta ({\bf r}')} = w({\bf r} - {\bf r}')
\end{equation}
In order to start with an action $S$ which is scale invariant, we are going to focus on the effects of short-ranged correlated disorder, taking in what follows the distribution $w({\bf r}) = w_0/{\bf r}^2$.

We consider the case of quenched disorder, which can be treated using the replica method. This leads us to introduce a number $n$ of different copies of the fields $\phi $ and $\psi_i$ in the action (\ref{act}), interacting with the random potential $\eta ({\bf r})$. Those fields get then an additional index $a$ running over the $n$ copies, with the need to take the limit $n \rightarrow 0$ at the end of every calculation in order to obtain physical quantities. 

\begin{figure}[h]
\begin{center}
\includegraphics[width=5.0cm]{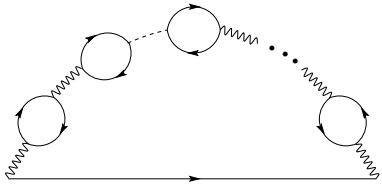}
\end{center}
\caption{Class of self-energy diagrams correcting the fermion propagator (represented by a directed line) to leading order in the large-$N$ limit, with multiple interactions mediated by the Coulomb potential (wavy line) and constrained to a single correlation of the disorder potential (dashed line) in order to render a nonvanishing contribution in the limit of number of replicas $n \rightarrow 0$.}
\label{one}
\end{figure}

The replica method can be applied to the computation of the electron self-energy in the large-$N$ limit. On the one hand, there is the RPA contribution of the disorder-free theory, already considered in Ref. \cite{rc}. On the other hand, the disorder potential introduces a new contribution, which is given in the large-$N$ limit by the sum of rainbow diagrams of the type shown in Fig. \ref{one}. Note that, in this approximation (which amounts also to neglect vertex corrections), the sum is restricted to diagrams with just a single correlation $w({\bf r})$, since the inclusion of more propagators of the random potential makes the corresponding rainbow diagram to vanish in the limit $n \rightarrow 0$.

The sum of the diagrams of the type shown in Fig. \ref{one} leads to a contribution to the electron self-energy
\begin{equation}
\Sigma_w ({\bf k},\omega ) =  \frac{w_0}{e_0^4} 
       \int \frac{d^3 p}{(2\pi )^3} \frac{2\pi^2}{|{\bf p}|} 
                \frac{G_0({\bf k}-{\bf p}, \omega ) }
        { \left(  \frac{1}{e_0^2} + \frac{N}{6\pi^2 v_F} \log \frac{\Lambda}{v_F |{\bf p}|}   \right)^2   }
\label{selft}
\end{equation}
where $G_0({\bf k}, \omega )$ stands for the free fermion propagator and $\Lambda $ is a high-energy cutoff. This has to be reabsorbed into a suitable redefinition of the parameters of the model to obtain meaningful physical results. Interestingly, that can be achieved by means of the same renormalization demanded by the disorder-free interacting theory, that is, by introducing the renormalized charge $e$ through the relation $1/e^2 = 1/e_0^2 + (N/6\pi^2 v_F) \log (\Lambda /\mu)$. To guarantee the cutoff independence of (\ref{selft}), one still has to impose that the prefactor $w_0/e_0^4$ does not depend on $\Lambda$, which can be done by introducing a renormalized disorder strength $w_R$ such that $w_R/e^4 = w_0/e_0^4$. Overall this implies the scaling of the couplings with the energy variable $\mu $
\begin{equation}
e^2 (\mu) = \frac{e_0^2}{1 + \frac{N e_0^2}{6\pi^2 v_F} \log \frac{\Lambda}{\mu}  }  \; , \; 
    w_R (\mu) = \frac{w_0}{\left( 1 + \frac{N e_0^2}{6\pi^2 v_F} \log \frac{\Lambda}{\mu} \right)^2 }
\label{renor}
\end{equation}
The last expression in (\ref{renor}) is just the reflection that the disorder is a marginally irrelevant perturbation of the interacting theory in the low-energy limit $\mu \rightarrow 0$.

The fact that the renormalized strength $w_R$ scales to zero at low energies does not mean however that there cannot be interesting effects driven by the disorder. In the present approach, those effects arise from the scaling of the electron quasiparticle parameters. The dressed fermion propagator $G({\bf k}, \omega )$ has to be independent of the high-energy cutoff $\Lambda $, which requires to introduce renormalization factors $Z_\psi$ for the quasiparticle weight and $Z_v$ for the Fermi velocity in the expression
\begin{equation}
G ({\bf k},\omega )^{-1}   =  Z_\psi (\omega  -  Z_v v_R  \gamma_0 \mbox{\boldmath $\gamma $}  \cdot  {\bf k}  ) - Z_\psi \Sigma ({\bf k},\omega )    
\end{equation}
In this last equation, $\Sigma $ stands for the sum of the disorder-free and the $\Sigma_w $ contribution. This latter becomes particularly important since it may lead to a significant reduction of the Fermi velocity, as we see in what follows.

In order to compute $Z_\psi$ and $Z_v$ up to high orders in $e^2$, it is convenient to turn to a dimensional regularization, instead of using the high-energy cutoff $\Lambda $. With that procedure, the momentum integrals are computed in dimension $D = 3 - \epsilon $, in such a way that the powers of $\log (\Lambda )$ are traded by poles in the $\epsilon $ parameter. In terms of the effective coupling $g = N e^2/2\pi^2 v_R$, the renormalization factors have in general the pole structure $Z_\psi = 1 + (1/N)\sum_{n=1}^{\infty } c_n (g)/\epsilon^n , Z_v = 1 + (1/N)\sum_{n=1}^{\infty } b_n (g)/\epsilon^n$. The disorder-free contribution to the coefficients $c_n$ and $b_n$ has been already studied in Ref. \cite{rc}. We report here the results from switching on the effects of disorder represented by the self-energy contribution $\Sigma_w $.

We note that the knowledge of $Z_\psi$ allows one to compute the electron anomalous dimension $\gamma_d $ from the dependence of the renormalized theory on the auxiliary energy scale $\mu$, which gives $\gamma_d (g) = (\mu /Z_\psi ) (\partial Z_\psi /\partial \mu )$ \cite{amit}.
Moreover, one can also exploit the fact that the unrenormalized theory does not know about $\mu $ to enforce the independence of the bare Fermi velocity on that energy scale, expressed as $\partial (Z_v v_R) / \partial \mu = 0$. This leads to a scaling equation for the renormalized Fermi velocity
\begin{equation}
\frac{\mu }{v_R }  \frac{\partial v_R}{\partial \mu }  = \beta (g)
\label{beta}
\end{equation}

\begin{figure}[h]
\begin{center}
\includegraphics[width=4.1cm]{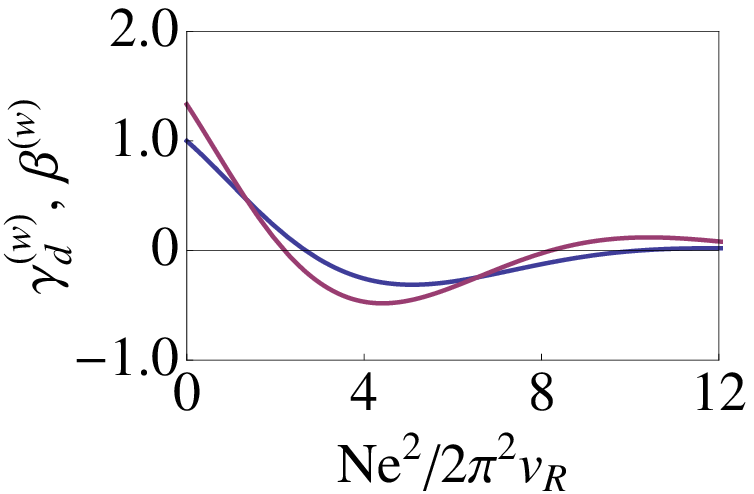}
\hspace{0.2cm}
\includegraphics[width=4.1cm]{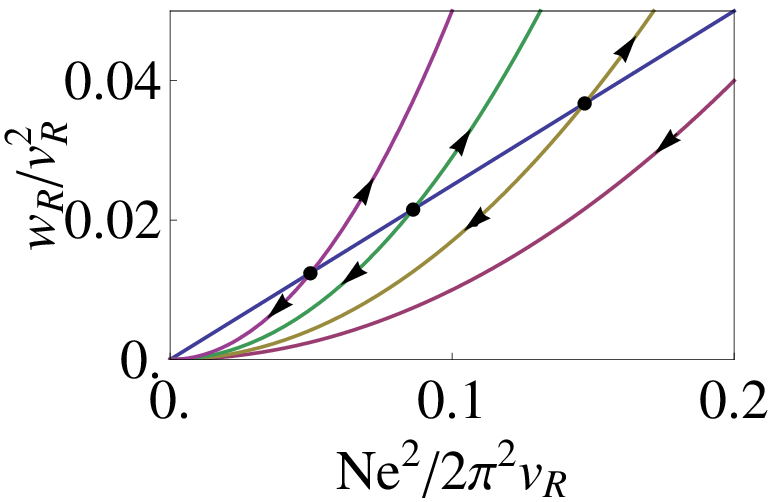}\\
 \hspace{0.36cm}  (a) \hspace{3.6cm} (b)
\end{center}
\caption{(a) Plot of the contribution from disorder to the electron anomalous dimension $\gamma_d $ (blue line) and to the rate of variation of the Fermi velocity $\beta $ (red line) in the large-$N$ approximation. (b) Renormalization group flow in the low-energy limit of the effective strengths of the disorder and the Coulomb interaction, obtained from the resolution of Eqs. (\ref{flow1}) and (\ref{flow2}). The blue line represents the set of unstable fixed-points at $w_R/v_R^2 = g/4$.}
\label{two}
\end{figure}

In practice, one takes advantage of the fact that only the first residues $c_1$ and $b_1$ contribute to $\gamma_d $ and $\beta $ \cite{ram}. We have computed in particular the part linear in $w_R$ of the residues $c_1$ and $b_1$ up to very large orders in the effective interaction strength $g = N e^2/2\pi^2 v_R$. These expansions, together with those in Ref. \cite{rc}, allow us to construct the electron anomalous dimension as $\gamma_d = (1/N) \gamma_d^{(0)} (g) + (w_R/v_R^2) \gamma_d^{(w)} (g) + O(w_R^2)$ and the scaling of the Fermi velocity as $\beta = (1/N) \beta^{(0)} (g) + (w_R/v_R^2) \beta^{(w)} (g) + O(w_R^2)$. The results obtained for $\gamma_d^{(w)} (g)$ and $\beta^{(w)} (g) $, valid up to values of $g$ deep into the strong-coupling regime, are plotted in Fig. \ref{two}(a).

It is important to stress that $\gamma_d^{(w)} (g) $ does not show any singular behavior in the range of couplings covered in Fig. \ref{two}(a). One of the main results reported in Ref. \cite{rc} was that the disorder-free contribution $\gamma_d^{(0)} (g)$ diverges at a critical value of the effective coupling $g_c = 3$. We may conclude therefore that the disorder effects here analyzed do not prevent the development of the non-Fermi liquid phase characteristic of the strong-interaction regime of Weyl semimetals. 

On the other hand, $\beta^{(w)}$ leads to a positive contribution to the right-hand-side of Eq. (\ref{beta}), which reads to lowest order in the couplings
\begin{equation}
\frac{\mu }{v_R }  \frac{\partial v_R}{\partial \mu }  = - \frac{1}{6\pi^2 } \frac{e^2}{v_R} 
             +  \frac{4}{3}  \frac{w_R}{v_R^2} +  \ldots
\end{equation}
This equation can be used to find the low-energy behavior of the effective strengths $g = N e^2/2\pi^2 v_R$ and $w_R/v_R^2$. The scaling in Eq. (\ref{renor}) can be encoded in the two equations  $\mu (\partial /\partial \mu ) e^2 = N e^4/6\pi^2 v_R $ and $\mu (\partial /\partial \mu ) w_R = N w_R e^2/3\pi^2 v_R $. Then, we get to quadratic order in the large-$N$ limit
\begin{eqnarray}
\mu \frac{\partial }{\partial \mu } g   & = &  \frac{1}{3} g^2 
             -  \frac{4}{3} \frac{w_R}{v_R^2} g   + \ldots         \label{flow1}         \\
\mu \frac{\partial }{\partial \mu } \frac{w_R}{v_R^2}   & = &  \frac{2}{3} \frac{w_R}{v_R^2} g 
             -  \frac{8}{3}  \left( \frac{w_R}{v_R^2} \right)^2     + \ldots 
\label{flow2}
\end{eqnarray}

Quite remarkably, Eqs. (\ref{flow1})-(\ref{flow2}) reveal the existence of a line of unstable fixed-points at $w_R/v_R^2 = g/4$, as shown in Fig. \ref{two}(b). Below that line, the theory scales in the low-energy limit $\mu \rightarrow 0$ towards the noninteracting regime. Above the critical line, the effective couplings flow away from the weak-coupling regime as a manifestation of the dominant effects of disorder, signaling the onset of a phase whose precise characterization requires a full nonperturbative approach.

{\em Schwinger-Dyson equations.---}
To get more information about the phase dominated by disorder, we resort to a self-consistent resolution of the Schwinger-Dyson equations of the electron system. In this approach, we are going to adopt a truncation of the equations that amounts to include all kind of diagrammatic contributions except those containing vertex corrections. Then, the electron propagator $G({\bf k}, \omega)$ is bound to satisfy the equation
\begin{equation}
i \Sigma ({\bf k},\omega ) = -\int \frac{d^3 p}{(2\pi )^3} \frac{d\omega_p }{2\pi } G({\bf k}-{\bf p}, \omega - \omega_p)  D({\bf p},\omega_p )
\label{si}
\end{equation}
where $D({\bf p},\omega_p )$ stands for the dressed interaction propagator. In the bare vertex approximation, that interaction includes the RPA sum of diagrams of the disorder-free theory plus a similar sum with just a single correlation of the disorder potential replacing each time one Coulomb potential (in analogy with the large-$N$ diagrams in Fig. \ref{one}). The difference with respect to the previous large-$N$ approach is that now the electron-hole polarization $\Pi ({\bf q},\omega_q )$ must be computed in terms of the dressed electron propagator according to the equation
\begin{equation}
i\Pi ({\bf q},\omega_q ) =  \int \frac{d^3 p}{(2\pi )^3} \frac{d\omega_p }{2\pi } {\rm Tr} \left[ G({\bf q}-{\bf p}, \omega_q - \omega_p)  G({\bf p},\omega_p )  \right]
\label{pi}
\end{equation}

It can be shown that the couple of equations (\ref{si})-(\ref{pi}) can be solved self-consistently by introducing the ansatz
\begin{equation}
G({\bf k}, \omega)  =  \left[  z_\psi({\bf k}, \omega) ( \omega  -  z_v({\bf k}, \omega) v_F \gamma_0 \mbox{\boldmath $\gamma $}  \cdot  {\bf k} ) \right]^{-1}
\label{prop}
\end{equation}
In practice, one may perform a numerical resolution of the integral equations by rotating all the frequencies in the complex plane, $\omega = i\overline{\omega} $, and working (in the case of the fermion propagator) with the discrete set $\overline{\omega}_n = (2n + 1)\pi k_B T$ running over integer numbers $n$. This amounts to place the theory at a finite temperature $T$. Furthermore, one has to cut off the integrals at a maximum value of the modulus of the momentum $\Lambda_k $ \cite{cut}. This momentum cutoff can be used then to assign the microscopic unrenormalized values of the physical parameters, like the bare Fermi velocity (as $v_B = z_v(\Lambda_k, 0) \: v_F$) or the bare electron charge (as $e_B^2 = \Lambda_k^2 \: D(\Lambda_k,0)$). 

\begin{figure}[h]
\begin{center}
\includegraphics[width=4.1cm]{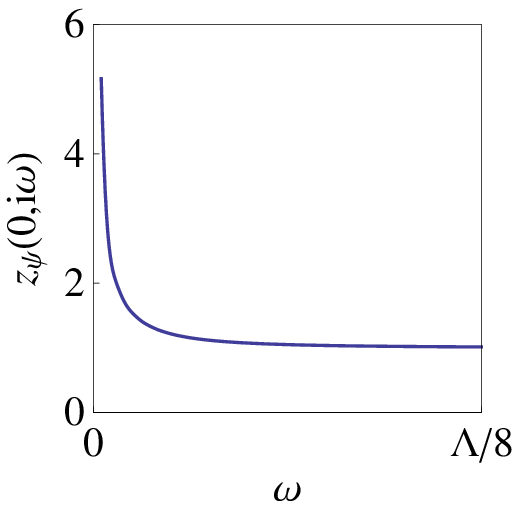}
\hspace{0.2cm}
\includegraphics[width=4.1cm]{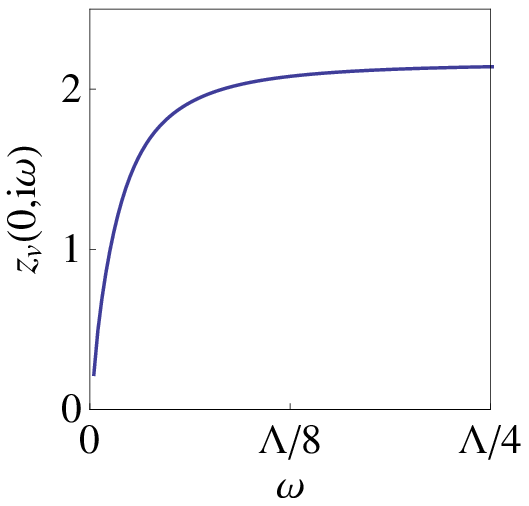}\\
 \hspace{0.36cm}  (a) \hspace{3.6cm} (b)
\end{center}
\caption{Renormalization factors $z_\psi $ (a) and $z_v $ (b) obtained from the self-consistent resolution of the Schwinger-Dyson equations for $N = 6$ and $w_0/v_B^2 \approx 17.0$, close to the disorder-driven transition (blue line) in the diagram of Fig. \ref{four}(a).}
\label{three}
\end{figure}

The resolution of the Schwinger-Dyson equations shows indeed that the model has two different phases (apart from the non-Fermi liquid phase at strong interaction identified in Ref. \cite{rc}) depending on the values of $w_0/v_B^2$ and $\lambda = e_B^2/2\pi^2 v_B$. It turns out that there is a critical line, in the regime of small $\lambda $, separating a phase with regular behavior of $z_\psi $ and $z_v $ from a different phase whose onset is characterized by the divergence of $z_\psi $ and concomitant vanishing of $z_v $ in the limit $\overline{\omega} \rightarrow 0$. This can be seen in Fig. \ref{three}, which shows a plot of those functions when the couplings reach the critical line.

\begin{figure}[h]
\begin{center}
\includegraphics[width=4.1cm]{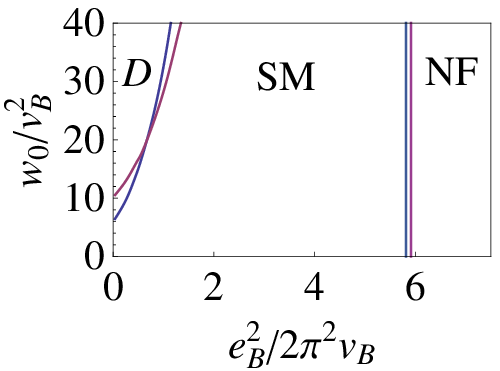}
\hspace{0.2cm}
\includegraphics[width=4.1cm]{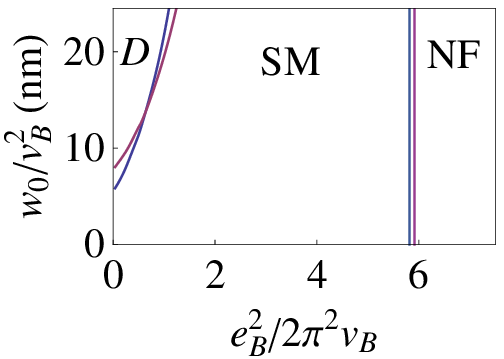}\\
 \hspace{0.36cm}  (a) \hspace{3.6cm} (b)
\end{center}
\caption{Phase diagrams of the interacting 3D Weyl semimetal (for $N = 6$) with (a) short-ranged-correlated and (b) uncorrelated disorder, showing the critical boundaries at $k_B T \approx 32$ meV (red lines) and $k_B T \approx 16$ meV (blue lines) which separate phases corresponding to a diffusive metal (D), semimetallic behavior (SM) and non-Fermi liquid behavior (NF).}
\label{four}
\end{figure}

The complete phase diagram as a function of the bare coupling strengths $w_0/v_B^2$ and $\lambda = e_B^2/2\pi^2 v_B$ is represented in Fig. \ref{four}(a) for the model with $N = 6$. It can be checked that the critical line separating the two phases at small $\lambda $ tends to reach the origin in the limit $T \rightarrow 0$, in accordance with the previous renormalization group results. Anyhow, the most important feature revealed by the present approach is that the phase induced by disorder disappears in the regime of intermediate interaction strength, which can be seen as a reflection of the prevalence of the screening effects in the interacting theory.

The present approach also allows us to uncover the physical meaning of the phase above the disorder-driven transition. The expression of the propagator (\ref{prop}) can be applied to compute the density of states $n(\omega )$ at the nodal points as
\begin{eqnarray}
n(0)  & = &  \lim_{\omega \rightarrow 0}  {\rm Im} \: {\rm Tr} \int d^3 k \: G({\bf k}, \omega)    \nonumber  \\
    & \sim &   \lim_{\omega \rightarrow 0}  {\rm Im} \: {\rm Tr} \int d^3 k \: \frac{1}{z_\psi z_v^3} G_0 ({\bf k}, \omega)   
            \sim  \lim_{\omega \rightarrow 0}  \frac{\omega^2 }{z_\psi z_v^3 }    \;\;\;\;\;
\end{eqnarray}
With this estimate, one can check that the vanishing of $z_v({\bf k}, 0) $ at the onset of the phase with dominant disorder corresponds to the appearance of a nonvanishing (not exponentially small) density of states in the limit $\omega \rightarrow 0$. We see therefore that the phase placed to the left in the diagram of Fig. \ref{four}(a) is similar to the diffusive metal phase induced by disorder in the noninteracting 3D semimetals\cite{fra,mura,gos,nomu,herb,brou,kosh,nand,radz,das,das2}. We can view then the critical line we have found as the extension of the disorder-driven transition of those systems, promoted here to a very steep boundary in the phase diagram which reflects the irrelevance of the disorder effects in the presence of a sufficiently strong Coulomb interaction.

{\em Conclusion.---}
In this paper we have studied the effect of a random disorder potential on interacting 3D Weyl semimetals, showing that the phase diagram of these systems contains in general three different phases. At strong interaction strength, we have seen that there is a non-Fermi liquid phase which corresponds to that identified in the disorder-free semimetal in Ref. \cite{rc}. In the weakly interacting theory, we have found a phase induced by disorder and characterized by the vanishing of the renormalized Fermi velocity at the nodal points, which is the analogue of the diffusive metal phase in the noninteracting semimetals. The resolution of the Schwinger-Dyson equations has also allowed us to show that the rest of the phase diagram is covered by a semimetallic phase with renormalized quasiparticle parameters, which extends over the whole regime of intermediate interaction strength.

Although we have dealt with a particular short-ranged-correlated disorder potential, we have checked that similar phases are obtained when the disorder is described by means of a random potential with delta-function correlation. In that case, a self-consistent resolution like that reported above leads to the phase diagram shown in Fig. \ref{four}(b). This corresponds to temperature $T \neq 0$, but the extrapolation of the results implies that the critical boundary to the left in Fig. \ref{four}(b) does not tend to reach the origin as $T \rightarrow 0$. This is the only qualitative difference with respect to the phases in Fig. \ref{four}(a), agreeing with a nonvanishing critical disorder strength for the transition driven by uncorrelated disorder in the noninteracting theory. 

The main practical conclusion of our work is that the effects of disorder can be in general disregarded in real 3D semimetals, due to the renormalization induced by the Coulomb interaction. For typical 3D semimetals with Fermi velocity $v_F \lesssim 1$ eV nm, the effective coupling $\lambda = e_B^2/2\pi^2 v_B$ gets values of order $\gtrsim 1$. This means that these systems should naturally fall in the regimes with intermediate or strong interaction strength, displaying semimetallic or non-Fermi liquid behavior, but away from the diffusive phase confined by the steep disorder-driven transition shown in Fig. \ref{four}.

\vspace{0.5cm}

{\em Acknowledgements.}
We acknowledge financial support from MINECO (Spain) through grant No. FIS2014-57432-P.


\begin{thebibliography}{24}
\bibitem{cut}
In our computation, the sum over frequencies is always cut off at a larger 
energy $\Lambda \gg v_F \Lambda_k $.
\end{thebibliography}
\end{document}